\documentclass[preprint,showpacs,preprintnumbers,amsmath,amssymb]{revtex4}

\newcommand{\cP}{\ensuremath{\mathcal{P}}}

\newcommand{\half}{{\textstyle\frac{1}{2}}}
\newcommand{\quarter}{{\textstyle\frac{1}{4}}}
\newcommand{\e}{{\rm e}} 

\begin{document}
\title{Wilson Polynomials and the Lorentz Transformation Properties of the
Parity Operator}

\author{Carl M.\ Bender, Peter N.\ Meisinger, and Qinghai Wang}

\affiliation{Department of Physics, Washington University, St.\ Louis, 
MO 63130, USA}

\date{\today}

\begin{abstract}
The parity operator for a parity-symmetric quantum field theory transforms as an
infinite sum of irreducible representations of the homogeneous Lorentz group.
These representations are connected with Wilson polynomials.
\end{abstract}
\pacs{11.30.Er, 11.30.Cp, 02.10.Nj, 02.20.-a}
\maketitle

\vskip2pc

\section{Introduction}
\label{s1}
The parity operator $\cP$ for a quantum field theory has the effect
$$\cP\varphi({\bf x},t)\cP=\varphi(-{\bf x},t)\quad({\rm scalar~field}),$$
$$\cP\varphi({\bf x},t)\cP=-\varphi(-{\bf x},t)\quad({\rm
pseudoscalar~field}).$$
We can also define an {\it intrinsic} parity operator
$\cP_I$. This operator has the same effect as $\cP$ except that it does not
reverse the sign of the spatial argument of the quantum field:
$$\cP_I\varphi({\bf x},t)\cP_I=\varphi({\bf x},t)\quad({\rm scalar~field}),$$
$$\cP_I\varphi({\bf x},t)\cP_I=-\varphi({\bf x},t)\quad({\rm
pseudoscalar~field}).$$

To determine the Lorentz transformation properties of an operator in a quantum
field theory one must calculate the commutator of this operator with the
generators of the Lorentz group $J^{\mu\nu}$. If this operator does not depend
on the space-time coordinates $({\bf x},t)$ and it commutes with $J^{\mu\nu}$,
then it is a scalar under Lorentz transformations.

For the purposes of this paper we will suppose that the Hamiltonian $H$ of our
quantum field theory has parity symmetry; that is, $[\cP,H]=0$. An example of
such a theory is the free scalar quantum field theory whose Hamiltonian is
$$H=\int d{\bf x}\,\left\{\half\pi^2({\bf x},t)+\half\left[\nabla\varphi({\bf x}
,t)\right]^2+\half\mu^2\varphi^2({\bf x},t)\right\}.$$
For such a theory it is easy to see that the intrinsic parity operator is a
scalar because 
$$[\cP_I,J^{\mu\nu}]=0.$$

However, while the conventional parity operator $\cP$ is a rotational scalar
because it commutes with the three generators of spatial rotations,
$$\left[{\cal P},J^{ij}\right]=0,$$
it does {\it not} commute with the Lorentz-boost generators:
$$\left[{\cal P},J^{0i}\right]=-2J^{0i}{\cal P}.$$
Therefore, the parity operator $\cP$ is not a Lorentz scalar. Furthermore, as we
will see, $\cP$ is not the spin-$0$ component of a vector, a tensor, or indeed
any finite-dimensional representation of the Lorentz group.

This observation raises a fundamental question: How does $\cP$ transform under
the Lorentz group? In this paper we provide a partial answer to this question.
We argue that $\cP$ transforms as an infinite direct sum of finite-dimensional
tensorial representations of the Lorentz group.

This paper is organized as follows: In Sec.~\ref{s2} we review the general
theory of the irreducible representations of the Lorentz group. Then in
Sec.~\ref{s3} we show that $\cP$ transforms as a direct sum of irreducible
representations of the Lorentz group, and that to identify each of these
irreducible representations one must solve a difference-equation eigenvalue
problem. In Secs.~\ref{s4} and \ref{s5} we perform an analysis of special cases
of this difference equation and show that the eigenfunctions are closely
associated with the Wilson polynomials. Finally, in Sec.~\ref{s6} we make some
concluding remarks about the structure of the solutions to the general
difference-equation eigenvalue problem that is derived in Sec.~\ref{s3}. The
properties of the Wilson polynomials are summarized in the Appendices.

\section{Irreducible Representations of the Lorentz Group}
\label{s2}
In this section we review briefly the classification of irreducible
representations of the Lorentz group. Following the exposition of the
irreducible representations of the Lorentz group by Gel'fand, Minlos, and
Shapiro \cite{Gel}, Bender and Griffiths \cite{BG}, and Griffiths \cite{G}, we
note that all irreducible representations of the Lorentz group are characterized
and labeled by a pair of numbers $\left(\ell_0,\ell_1\right)$. The first number,
$\ell_0$, is a nonnegative integer and the second number, $\ell_1$, is in
general complex. Any irreducible representation of the Lorentz group is a direct
sum of irreducible representations of the rotation subgroup $SO(3)$ and $\ell_0$
is the lowest-spin of the representations contained in this sum.

If $\ell_1-\ell_0 =0$ or noninteger, then the representation belongs to the {\it
nonsingular class} of infinite-dimensional representations. Such a
representation contains an infinite tower of spins beginning with $\ell_0$; the
spins $\ell$ in this representation are $\ell=\ell_0,~\ell_0+1,~\ell_0+2,~
\cdots$. Each spin in this sequence occurs once and only once.

If $\ell_1-\ell_0$ is a nonzero integer, then the representation belongs to the
{\it singular class}. For this case if $|\ell_1|>\ell_0$, the representation is 
finite-dimensional; it contains all spins $\ell$ with $\ell_0 \leq \ell \leq
|\ell_1|-1$ and each spin occurs exactly once. If $|\ell_1|<\ell_0$, the
representation is infinite-dimensional; it contains all spins $\ell$: $\ell_0
\leq\ell<\infty$ and again each spin occurs exactly once.

Since the Lorentz group is noncompact, it also has indecomposable
representations. However, this situation does not arise in this paper.

To describe the Lorentz transformation properties of a space-time-dependent
quantity, such as a quantum field, we must specify all of the spin components of
the field $\left(Q_{\ell_0},~Q_{\ell_0+1},~Q_{\ell_0+2},~\cdots\right)$. Let us
suppose first that this field transforms irreducibly under the Lorentz group and
let us also suppose that the lowest-spin component has spin $0$. Then, an
infinitesimal Lorentz transformation of the spin-$0$ component $Q_0({\bf x},t)$
has an orbital and a spin part. Specifically,
\begin{equation}
-i\left[Q_0,J^{0i}\right]=t\partial^i Q_0-x^i\partial^0 Q_0+Q_1^i,
\label{e1}
\end{equation}
where $Q_1^i$ is the spin-$1$ component of the irreducible representation. Under
an infinitesimal Lorentz transformation, this spin-$1$ component then transforms
like
\begin{equation}
-i\left[Q_1^j,J^{0i}\right]=t\partial^i Q_1^j-x^i\partial^0 Q_1^j+Q_2^{
ij}+\alpha\delta^{ij}Q_0,
\label{e2}
\end{equation}
where $Q_2^{ij}$ is the spin-$2$ component of the representation. Note that
$Q_2^{ij}$ is a {\it pure} spin-$2$ object and is therefore symmetric and
traceless: $Q_2^{ii}=0$. (Repeated indices indicate summation.) The number
$\alpha$ in (\ref{e2}) is directly related to the parameter $\ell_1$ for the
irreducible representation by the formula \cite{BG}
\begin{equation}
\alpha={\textstyle\frac{1}{3}}\left(\ell_1^2-1\right).
\label{e3}
\end{equation}
This process of evaluating commutators can be used iteratively to calculate all
of the spin components of the irreducible representation.

Now suppose that $Q_0$ is the spin-$0$ component of a representation of the
Lorentz group, but that this representation is not irreducible. This means that
there may be more than one component of each spin. Thus, under an infinitesimal
transformation we have
\begin{equation}
-i\left[Q_0,J^{0i}\right]=t\partial^iQ_0-x^i\partial^0Q_0+N_1^i(Q_0),
\label{e4}
\end{equation}
where $N_1^i$ is one of the spin-$1$ components of the representation. An
infinitesimal Lorentz transformation of this spin-$1$ component produces a
spin-$2$ component $N_2^{ij}(Q_0)$, for which $N_2^{ii}(Q_0)=0$, and a new
spin-$0$ component $N_0(Q_0)$:
\begin{equation}
-i\left[N_1^j(Q_0),J^{0i}\right]=t\partial^iN_1^j(Q_0)-x^i\partial^0
N_1^j(Q_0)+N_2^{ij}(Q_0)+\delta^{ij}N_0(Q_0).
\label{e5}
\end{equation}
Since this representation is not irreducible, $N_0(Q_0)$ is not a multiple of
$Q_0$.

This process of calculating commutators can again be used to generate all of the
spin components of the representation. For example, commuting $N_2^{ij}(Q_0)$
with the Lorentz boost will define a spin-$3$ component $N_3^{ijk}(Q_0)$.
However, the procedure is significantly more complicated than that for an
irreducible representation because there may be many components for each spin.
Indeed, it may be that there are an infinite number of spin-$0$ components,
$Q_0$, $N_0(Q_0)$, $N_0'(Q_0)$, and so on, and we will see that this is the case
with the parity operator $\cP$.

\section{Infinitesimal Lorentz transformation of the $\cP$ Operator}
\label{s3}
Like the generators of space-time translations $P^\mu$, the parity operator
$\cP$ is a global symmetry operator and is not an explicit function of the
space-time coordinates $x^\mu$; that is, $\partial^\mu{\cal P}=0$. As a
consequence, the orbital terms in (\ref{e4}) vanish and we have
\begin{equation}
N_1^i({\cal P})\equiv-i[{\cal P},J^{0i}]=2iJ^{0i}{\cal P}.
\label{e6}
\end{equation}
This identifies the spin-$1$ term that arises after commuting with the generator
of a Lorentz boost.

A second commutation with the Lorentz boost gives a spin-$2$ component and a
new spin-$0$ component:
\begin{equation}
-i[N_1^j({\cal P}),J^{0i}]=N_2^{ij}({\cal P})+\delta^{ij}N_0({\cal P}),
\label{e7}
\end{equation}
where the new spin-$0$ component is
\begin{equation}
N_0({\cal P})={\textstyle\frac{4}{3}}J^{0i}J^{0i}{\cal P}.
\label{e8}
\end{equation}
Since $N_0({\cal P})$ is not a constant multiple of $\cP$, we conclude
immediately that $\cP$ does not belong to an irreducible representation of
the Lorentz group.

To analyze the Lorentz transformation properties of the parity operator it is
convenient to introduce the following notation. First, we define two
$3$-vectors:
\begin{eqnarray}
K^i &\equiv& J^{0i},\nonumber\\
L^i &\equiv& \half\epsilon^{ijk}J^{jk}.
\label{e9}
\end{eqnarray}
These operator quantities satisfy the following commutation relations:
\begin{eqnarray}
[L^i,L^j] &=& i\epsilon^{ijk}L^k,\nonumber\\
\left[K^i,L^j\right] &=& i\epsilon^{ijk}K^k,\nonumber\\
\left[K^i,K^j\right] &=& -i\epsilon^{ijk}L^k.
\label{e10}
\end{eqnarray}
Next, we define three operator products:
\begin{eqnarray}
W &\equiv& K^iK^i,\nonumber\\
A &\equiv& L^iL^i,\nonumber\\
{\rm m} &\equiv& K^iL^i=L^i K^i.
\label{e11}
\end{eqnarray}
These three operators are mutually commuting:
\begin{equation}
[W,A]=[A,{\rm m}]=[{\rm m},W]=0.
\label{e12}
\end{equation}
The $W$, $A$, and ${\rm m}$ operators have the following commutation
relations with the generators of the Lorentz group:
\begin{eqnarray}
\left[W,K^i\right]=[A,K^i]=-2i\epsilon^{ijk}L^jK^k-2K^i,
\nonumber\\
\left[W,L^i\right]=[A,L^i]=[{\rm m},K^i]=[{\rm m},L^i]=0.
\label{e13}
\end{eqnarray}

Next, we define the operator $B$:
\begin{equation}
B\equiv A-W.
\label{e14}
\end{equation}
Observe that from (\ref{e13}) the operator $B$ commutes with the generators
of the Lorentz group:
\begin{equation}
[B,K^i]=[B,L^i]=0.
\label{e15}
\end{equation}
Also, $B$ commutes with the parity operator ${\cal P}$:
\begin{equation}
[B,{\cal P}]=0.
\label{e16}
\end{equation}
The vanishing of the commutation relations in (\ref{e15}) and (\ref{e16})
imply that $B$ transforms as a scalar. The operator ${\rm m}$ also commutes with
the generators of the Lorentz group but ${\rm m}$ is a pseudoscalar because
\begin{equation}
{\cal P}{\rm m}{\cal P}=-{\rm m}.
\label{e17}
\end{equation}
Finally, we define the scalar operator $M$ by
\begin{equation}
M \equiv {\rm m}^2.
\label{e18}
\end{equation}
The four operators $B$, $M$, $W$, and parity $\cP$ are a mutually commuting set:
\begin{equation}
[B,M]=[B,W]=[M,W]=[B,\cP]=[M,\cP]=[W,\cP]=0.
\label{e19}
\end{equation}

We know that the parity operator does not transform irreducibly under the
Lorentz group. Thus, we will assume that it transforms as a direct sum of
operators that {\it do} transform irreducibly. From the commutation properties
of the operators defined above we know that the most general operator that could
be produced by repeated commutation of $\cP$ with the generators of the Lorentz
group has the form $f(B,M,W)\cP$, where $f(B,M,W)$ is an as yet unknown function
of the operators $B$, $M$, and $W$. We can see from (\ref{e8}) that this is the
most general spin-$0$ structure that can appear with repeated commutation. Let
us assume that the function $f(B,M,W)$ has a power series representation of the
form
\begin{equation}
f(B,M,W)=\sum_{l,m,n}a_{lmn}B^lM^mW^n.
\label{e20}
\end{equation}

A lengthy calculation of the double commutator of $f(B,M,W)\cP$ with the
generators of the Lorentz group gives the new spin-$0$ component
\begin{eqnarray}
N_0[f(B,M,W){\cal P}] &=& -\frac{2}{3}\left(W+\frac{M}{B+W}\right)f(B,M,W){\cal
P}-\frac{1}{3\sqrt{1+4B+4W}}\biggl\{\nonumber\\
&&\qquad\left[2B+4W+\left(W-\frac{M}{B+W}\right)\left(\sqrt{1+4B+4W}-1\right)
\right]\nonumber\\
&& \qquad\qquad\times\ f\left(B,M,W+1+\sqrt{1+4B+4W}\right)\nonumber\\
&& \qquad+\left[-2B-4W+\left(W-\frac{M}{B+W}\right) 
\left(\sqrt{1+4B+4W}+1\right)\right]\nonumber\\
&& \qquad\qquad\times f\left(B,M,W+1-\sqrt{1+4B+4W}\right)\biggr\}{\cal P}.
\label{e21}
\end{eqnarray}
Let us now suppose that $f(B,M,W){\cal P}$ transforms as the irreducible
$(0,\ell_1)$ representation of Lorentz group. Then, from (\ref{e3}) we have
\begin{equation}
N_0[f(B,M,W){\cal P}]={\textstyle\frac{1}{3}}(\ell_1^2-1)f(B,M,W){\cal P}.
\label{e22}
\end{equation}
Combining (\ref{e21}) and (\ref{e22}), we conclude that $f(B,M,W)$ satisfies the
functional relation
\begin{eqnarray}
&& 2\left(W+\frac{M}{B+W}\right)f(B,M,W)+\frac{1}{\sqrt{1+4B+4W}}\biggl\{
\nonumber\\
&&\qquad\left[2B+4W+\left(W-\frac{M}{B+W}\right)\left(\sqrt{1+4B+4W}-1\right)
\right]\nonumber\\
&& \qquad\qquad\times f\left(B,M,W+1+\sqrt{1+4B +4W}\right)\nonumber\\
&& \qquad+\left[-2B-4W+\left(W-\frac{M}{B+W}\right)\left(\sqrt{1+4B+4W}+1\right)
\right]\nonumber\\
&& \qquad\qquad\times f\left(B,M,W+1-\sqrt{1+4B+4W}\right)\biggr\}=(1-\ell_1^2)
f(B,M,W).
\label{e23}
\end{eqnarray}

This functional equation is a {\it difference-equation eigenvalue problem} in
which the eigenvalue $(1-\ell_1^2)$ determines the representation of the Lorentz
group under which the operator $f(B,M,W)\cP$ transforms. We do not yet know how
to find the general analytical solution to this equation. However, in the next
two sections we show how to solve analytically this master equation for two
important special cases.

\section{Special Case I: $B=0$ and $M=0$}
\label{s4}

In this section we solve an important special case of the difference-equation
eigenvalue problem (\ref{e23}). Let us assume that $B=0$ and $M=0$. Then,
(\ref{e23}) simplifies to
\begin{eqnarray}
&& 2Wf(W)+\frac{\left(3+\sqrt{1+4W}\right)W}{\sqrt{1+4W}}f\left(W+1+\sqrt{1+4W}
\right)\nonumber\\
&&\qquad -\frac{\left(3-\sqrt{1+4W}\right)W}{\sqrt{1+4W}}f\left(W+1-\sqrt{1+4W}
\right)=(1-\ell_1^2)f(W).
\label{e24}
\end{eqnarray}
To solve this equation we make the change of independent variable
\begin{equation}
z=\sqrt{W+\quarter}.
\label{e25}
\end{equation}
In terms of this new variable $f(W)= f\left(z^2-\quarter\right)$.
Next, we introduce the dependent variable $g(z)$ by
\begin{equation}
g(z)=\e^{-i\pi z}\frac{f\left(z^2-\quarter\right)}{z^2-\quarter}.
\label{e26}
\end{equation}
The functional equation satisfied by $g(z)$ is
\begin{eqnarray}
&&(2z+1)(2z+3)^2g(z+1)-4z(2z-1)(2z+1)g(z)+(2z-1)(2z-3)^2g(z-1)\nonumber\\
&=& 8z(\ell_1^2-1)g(z).
\label{e27}
\end{eqnarray}

In order to determine the eigenvalues $(\ell_1^2-1)$ we must impose an
eigenvalue condition. The condition that we will use is that $f(W)$ be an {\it
entire} function of $W$. This choice is a natural physical constraint because it
eliminates the possibility that the irreducible representation of the Lorentz
group in (\ref{e22}) could be a nonsmooth function of the physical parameters
$B$, $M$, and $W$. A similar eigenvalue condition based on the entirety of a
solution to a difference equation may be found in Ref.~\cite{BO}.

Imposing this constraint gives an infinite sequence of allowed values for
$\ell_1$:
\begin{equation}
\ell_1(n)=2n+1\quad(n=0,1,2,3,\cdots).
\label{e28}
\end{equation}
Note that the eigenvalues involve the square of $\ell_1$ but we may assume that
$\ell_1$ is positive \cite{POS}. The irreducible representations of the Lorentz
group corresponding to these values of $\ell_1(n)$ are traceless, totally
symmetric tensors of rank $2n$: $T$, $T^{\mu\nu}$, $T^{\mu\nu\lambda\sigma}$,
and so on. Thus, we see that the parity operator transforms as a direct sum of
the spin-$0$ components of these tensors [see (\ref{e46})].

The eigenfunctions corresponding to the above eigenvalues are all polynomials in
$z^2$ except for the eigenfunction corresponding to the lowest eigenvalue for
which the eigenfunction is a rational function:
\begin{eqnarray}
g_0(z) &=& \frac{1}{z^2-\quarter},\nonumber\\
g_1(z) &=& 1,\nonumber\\
g_2(z) &=& z^2+{\textstyle\frac{3}{4}},\nonumber\\
g_3(z) &=& z^4+{\textstyle\frac{7}{2}}z^2+{\textstyle\frac{117}{80}},\nonumber\\
g_4(z) &=& z^6+{\textstyle\frac{37}{4}}z^4+{\textstyle\frac{1957}{112}}z^2
+{\textstyle\frac{2385}{448}},\nonumber\\
g_5(z) &=& z^8+19z^6+{\textstyle\frac{747}{8}}z^4+{\textstyle\frac{2011}{16}}z^2
+{\textstyle\frac{55575}{1792}},
\label{e29}
\end{eqnarray}
and so on. We have normalized the eigenfunctions so that each of the polynomials
is {\it monic}; that is, the coefficient of the highest power of $z^2$ is unity.

The polynomial eigenfunctions in (\ref{e29}) are {\it Wilson} polynomials
${\rm W}_n(x^2;a,b,c,d)$, where the parameters are given by
$a=b=1/2$ and $c=d=3/2$ (see Appendix A):
\begin{equation}
g_n(z)=\frac{(n+1)!}{(2n)!}{\rm W}_{n-1}\left(-z^2;\half,\half,{\textstyle\frac{
3}{2}},{\textstyle\frac{3}{2}}\right).
\label{e30}
\end{equation}
These polynomials can be expressed as generalized hypergeometric functions:
\begin{equation}
g_n(z)=\frac{(n-1)!(n!)^2(n+1)!}{(2n)!}{}_4{\rm F}_3\left(1-n,n+2,
\half-z,\half+z;1,2,2;1\right).
\label{e31}
\end{equation}

From the solutions $g_0(z)$ in (\ref{e29}) and $g_n(z)$ in (\ref{e31}) we can
construct the eigenfunction solutions to (\ref{e24}):
\begin{equation}
f_n(W)= \begin{cases}
\e^{i\pi\sqrt{W+1/4}} ~~(n=0),\\
\e^{i\pi\sqrt{W+1/4}}\frac{(n-1)!(n!)^2(n+1)!}{(2n)!}W\\ ~~
\times~{}_4{\rm F}_3 \left(1-n,n+2,\half - {\sqrt{W+\quarter}},\half+
{\sqrt{W+\quarter}};1,2,2;1\right) ~~(n>0).
\end{cases}
\label{e32}
\end{equation}
Here are the first six solutions:
\begin{eqnarray}
f_0(W) &=& \e^{i\pi\sqrt{W+1/4}},\nonumber\\
f_1(W) &=& \e^{i\pi\sqrt{W+1/4}}W,\nonumber\\
f_2(W) &=& \e^{i\pi\sqrt{W+1/4}}W\left(W+1\right),\nonumber\\
f_3(W) &=& \e^{i\pi\sqrt{W+1/4}}W\left(W^2+4W+{\textstyle\frac{12}{5}}\right),
\nonumber\\
f_4(W) &=& \e^{i\pi\sqrt{W+1/4}}W\left(W^3+10W^2+{\textstyle\frac{156}{7}}W+
{\textstyle\frac{72}{7}}\right),\nonumber\\
f_5(W) &=& \e^{i\pi\sqrt{W+1/4}}W\left(W^4+20W^3+108W^2+176W+{\textstyle\frac{
480}{7}}\right).
\label{e33}
\end{eqnarray}

The difference equation (\ref{e27}) is second order and linear, and this means
that for each eigenvalue there are two linearly independent solutions. It is
straightforward to use the method of reduction of order (which is ordinarily
used for linear differential equations) to calculate the second solution. We
seek a second solution of the general form
\begin{equation}
h_n(z)=g_n(z)u_n(z),
\label{e34}
\end{equation}
where $u_n(z)$ is an unknown function to be determined. The function $u_n(z)$ is
easy to find because it satisfies a first-order difference equation:
\begin{equation}
\frac{u_n(z+1)-u_n(z)}{u_n(z)-u_n(z-1)}=\frac{(2z-1)(2z-3)^2g_n(z-1)}{(2z+1)
(2z+3)^2g_n(z+1)}.
\label{e35}
\end{equation}
From this equation we find that apart from a multiplicative constant, we have
\begin{equation}
u_n(z)-u_n(z-1)=\frac{1}{(2z-3)^2(2z-1)^3(2z+1)^2g_n(z-1)g_n(z)}.
\label{e36}
\end{equation}
Summing both sides of this equation, we get $u_n(z)$ (apart from an additive
constant):
\begin{equation}
u_n(z)=\sum_{x=z_0}^z\frac{1}{(2x-3)^2(2x-1)^3(2x +1)^2g_n(x-1)g_n(x)},
\label{e37}
\end{equation}
where $z_0$ is arbitrary.

This sum can be evaluated analytically for the case $n=0$ and we obtain
\begin{equation}
h_0(z)=\frac{1}{(z^2-\quarter)^2}.
\label{e38}
\end{equation}
When $n>0$ we must leave the second solution in the form of a sum:
\begin{equation}
h_n(z)=g_n(z)\sum_{x=z_0}^z\frac{1}{(2x-3)^2(2x-1)^3(2x+1)^2g_n(x-1)g_n(x)}.
\label{e39}
\end{equation}
The general solution to (\ref{e27}) is a linear combination of $g_n(z)$ and $h_n
(z)$. However, as a quantization condition, if we demand that the solution to
(\ref{e24}) be entire, then the coefficient of $h_n(z)$ must vanish.

Our objective now is to reconstruct the parity operator as a direct sum over the
irreducible representations constructed by multiplying $f_n(W)$ in (\ref{e32})
by the parity operator $\cP$:
\begin{equation}
{\cal P}=\sum_{n=0}^\infty c_n f_n(W){\cal P},
\label{e40}
\end{equation}
where $\{c_n\}$ are coefficients that must be determined from the equation
\begin{equation}
1=\sum_{n=0}^\infty c_nf_n(W)=\e^{i\pi z}\left[c_0+\left(z^2-\quarter\right)
\sum_{n=1}^\infty c_n g_n(z)\right],
\label{e41}
\end{equation}
which is obtained from (\ref{e40}) by multiplying by $\cP$.

Equation (\ref{e41}) must hold for every value of $z$. Thus, if we let
$z=\half$, we obtain
\begin{equation}
c_0=-i.
\label{e42}
\end{equation}
For $n\geq 1$, the coefficients $c_n$ satisfy
\begin{equation}
\frac{\e^{-i\pi z}+i}{z^2-\quarter}=\sum_{n=1}^\infty c_ng_n^{(1)}(z)=
\sum_{n=0}^\infty(-1)^n c_{n+1}P_n(-z^2),
\label{e43}
\end{equation}
where $P_n(x^2)$ is defined in (\ref{ea1}). We solve this formal functional
equation by continuing it analytically into the complex $z$-plane and then
making the change of variable $x=iz$. Equation (\ref{e43}) then becomes
\begin{equation}
-\frac{4\left(\e^{-x\pi}+i\right)}{1+4x^2}=\sum_{n=0}^\infty(-1)^nc_{n+1}P_n(x^2
).
\label{e44}
\end{equation}
We then multiply both sides of this equation by $\quarter\pi^2x(1+4x^2)^2\sinh(x
\pi)P_m(x^2)/\cosh^3(x\pi)$ and integrate over $x$ from $0$ to $\infty$. Using
the orthogonality property of the Wilson polynomials in (\ref{ea2}), we obtain
the following quadrature formula for $c_n$ for $n\geq1$:
\begin{equation}
c_n=\pi^2\frac{(-1)^n[(2n)!]^2(2n+1)!}{[(n-1)!]^2(n!)^4[(n+1)!]^4}\int_0^\infty
dx\,x(1+4x^2)\frac{\left(\e^{-x\pi}+i\right)\sinh(x\pi)}{\cosh^3(x\pi)}P_{n-1}(
x^2). 
\label{e45}
\end{equation}

Note that this analytic continuation is an extremely delicate process because
the series in (\ref{e40}) has only a formal existence. Indeed, while the Wilson
polynomials (\ref{ea1}) are complete and orthogonal, the polynomials obtained by
replacing $x^2$ by $-z^2$ [see (\ref{e43})] are not orthogonal with a positive
weight function and not complete in the usual sense. Thus, the analytic
continuation above is a procedure that converts a purely formal series identity
into a series that actually converges with coefficients that can be determined.

If we substitute the expression for $c_n$ ($n\geq1$) in (\ref{e45}) and also the
value of $c_0$ in (\ref{e42}) into (\ref{e40}), then (\ref{e40}) becomes an
identity. This implies that we have decomposed the representation under which
$\cP$ transforms into a direct sum of finite-dimensional irreducible
representations
\begin{equation}
(0,1)\oplus(0,3)\oplus(0,5)\oplus(0,7)\oplus\cdots.
\label{e46}
\end{equation}
That is, we have shown that $\cP$ transforms as a scalar plus the spin-$0$
component of a two-index symmetric traceless tensor plus the spin-$0$ component
of a four-index symmetric traceless tensor plus the spin-$0$ component of a
six-index symmetric traceless tensor, and so on. This is the central result
of our analysis.

We conclude this section with the remark that the first term in the direct sum
above is a scalar; that is, the term $f_0(W){\cal P}$ transforms like a Lorentz
scalar. Thus, $\cP$ transforms like the elementary structure $\e^{-i\pi\sqrt{W+
1/4}}$.

\section{Special Case II: $M=0$}
\label{s5}

In this section we solve an eigenvalue equation that is much more general than
that solved in Sec.~\ref{s4}. We consider here the case $M=0$ but we allow $B$ 
and $W$ to be arbitrary. Now, (\ref{e23}) simplifies to
\begin{eqnarray}
&& 2Wf(B,W)+\frac{1}{\sqrt{1+4B+4W}}\nonumber\\
&& \times\left[\left(2B+3W+W\sqrt{1+4B+4W}\right)f\left(B,W+1+\sqrt{1+4B+4W}
\right)\right.\nonumber\\
&&\quad\left.-\left(2B+3W-W\sqrt{1+4B+4W}\right) 
f\left(B,W+1-\sqrt{1+4B +4W}\right) \right]\nonumber\\
&=&(1-\ell_1^2)f(B,W).
\label{e47}
\end{eqnarray}

To analyze this equation we generalize the substitution that we made in
(\ref{e25}) and make the change of independent variable
\begin{equation}
z=\sqrt{W+B+\quarter}.
\label{e48}
\end{equation}
In terms of this new variable we have $f(B,W)=f\left(B,z^2-B-\quarter\right)$.
Next, we introduce the new dependent variable by the substitution
\begin{equation}
g(B,z)=\e^{-i\pi z}f\left(B,z^2-B-\quarter\right),
\label{e49}
\end{equation}
which is a generalization of (\ref{e26}). We obtain the following
functional equation satisfied by $g(B,z)$:
\begin{eqnarray}
&&2\left(z^2-B-\frac{1}{4}\right)g(B,z)-\frac{1}{2z}\left[3z^2-B-\frac{3}{4}+2z
\left(z^2-B-\frac{1}{4}\right) \right] g(B,z+1)\nonumber\\
&& \quad+\frac{1}{2z}\left[3z^2-B-\frac{3}{4}-2z\left(z^2-B-\frac{1}{4}\right)
\right]g(B,z-1)=\left(1-\ell_1^2\right)g(B,z).
\label{e50}
\end{eqnarray}

The solutions to this difference equation are Wilson polynomials ${\rm W}_n(x^2;
a,b,c,d)$ with parameters $a=b=\half$, $c=\half-\sqrt{B+1}$, $d=\half+
\sqrt{B+1}$ (see Appendix B). Thus,
\begin{eqnarray}
g_n(B,z) &=& \frac{n!}{(2n)!}{\rm W}_n\left(-z^2,\half,\half,
\half-\sqrt{B+1},\half+\sqrt{B+1}\right)\nonumber\\
&=& \frac{(n!)^2\Gamma\left(n+1-\sqrt{B+1}\right) 
\Gamma\left(n+1+\sqrt{B+1}\right)}{(2n)!\,\Gamma\left(1-\sqrt{B+1}\right) 
\Gamma\left(1+\sqrt{B+1}\right)}\nonumber\\
&&\quad\times~{}_4{\rm F}_3\left(-n,n+1,\half-z,\half+z;1, 
1-\sqrt{B+1},1+\sqrt{B+1};1\right).
\label{e51}
\end{eqnarray}
The first four of these polynomials are
\begin{eqnarray}
g_{0}(B,z) &=& 1,\nonumber\\
g_{1}(B,z) &=& z^2-\half B-\quarter,\nonumber\\
g_{2}(B,z) &=& z^4-\left(B-\half\right)z^2+{\textstyle\frac{1}{6}}B^2-\quarter B
-{\textstyle\frac{3}{16}},\nonumber\\
g_{3}(B,z) &=& z^6-\left({\textstyle\frac{3}{2}}B-{\textstyle\frac{13}{4}}
\right)z^4+\left({\textstyle\frac{3}{5}}B^2-{\textstyle\frac{57}{20}}B+
{\textstyle\frac{47}{80}}\right)z^2-{\textstyle\frac{1}{20}}B^3+{\textstyle
\frac{2}{5}}B^2-{\textstyle\frac{63}{160}}B-{\textstyle\frac{117}{320}}.
\label{e52}
\end{eqnarray}

Correspondingly, the solutions of (\ref{e47}) are
\begin{eqnarray}
f_n(B,W)&=&\e^{i\pi\sqrt{W+B+1/4}}\frac{(n!)^2\Gamma\left(n+1-\sqrt{B+1}\right) 
\Gamma\left(n+1+\sqrt{B+1}\right)}{(2n)!\,\Gamma\left(1-\sqrt{B+1}\right)\Gamma
\left(1+\sqrt{B+1}\right)}\nonumber\\
&&\hspace{-2.3cm}\times~{}_4{\rm F}_3\left(\!-n,n+1,\half-\sqrt{W\!+\!B\!+\!
\quarter},\half+\sqrt{W\!+\!B\!+\!\quarter};1,1-\sqrt{B\!+\!1},1+\sqrt{B\!+\!1};
1\!\right)
\label{e53}
\end{eqnarray}
and the first four of these solutions are
\begin{eqnarray}
f_0(B,W) &=& \e^{i\pi\sqrt{B+W+1/4}},\nonumber\\
f_1(B,W) &=& \e^{i\pi\sqrt{B+W+1/4}}\left(\half B+W\right),\nonumber\\
f_2(B,W) &=& \e^{i\pi\sqrt{B+W+1/4}}\left[{\textstyle\frac{1}{6}}B^2+B\left(W+
\half\right)+W(W+1)\right],\nonumber\\
f_3(B,W) &=& \e^{i\pi\sqrt{B+W+1/4}}\left[{\textstyle\frac{1}{20}}B^3+B^2\left(
{\textstyle\frac{3}{5}}W+{\textstyle\frac{19}{20}}\right)+B\left({\textstyle
\frac{3}{2}}W^2+{\textstyle\frac{22}{5}}W+{\textstyle\frac{6}{5}}\right)\right.
\nonumber\\
&&\qquad\qquad\left.+W\left(W^2+4W+{\textstyle\frac{12}{5}}\right)\right].
\label{e54}
\end{eqnarray}
It is interesting that while we are now treating $B$ as an arbitrary parameter,
the eigenvalues are independent of $B$ and thus are the same as in the case
considered in Sec.~\ref{s4} [see (\ref{e28})].

To find the Lorentz transformation properties of the parity operator we need to
reconstruct the parity operator in terms of these solutions $f_n(B,W)$:
\begin{equation}
{\cal P}=\sum_{n=0}^\infty c_nf_n(B,W){\cal P}.
\label{e55}
\end{equation}
This requires that we find the coefficients $c_n$ in the identity
\begin{equation}
1=\sum_{n=0}^\infty c_n f_n(B,W)=\e^{i\pi z}\sum_{n=0}^\infty c_ng_n(B,z).
\label{e56}
\end{equation}
To analyze this identity we multiply it by $\e^{-i\pi z}$ and obtain
\begin{equation}
\e^{-i\pi z}=\sum_{n=0}^\infty c_ng_n(B,z).
\label{e57}
\end{equation}
Next, we continue analytically this formal functional equation to the complex
$z$-plane and let $x=iz$. Equation (\ref{e57}) now becomes
\begin{equation}
\e^{-x\pi}=\sum_{n=0}^\infty(-1)^nc_{n}P_n(x^2),
\label{e58}
\end{equation}
where $P_n(x^2)$ is defined in (\ref{eb1}). Next, we multiply the both sides of
(\ref{e58}) by 
$$\frac{4\pi^2(2n)!\,(2n+1)!\,x\tanh(x\pi)}{(n!)^4\Gamma^2\left(n+1-\sqrt{B+1}
\right)\Gamma^2\left(n+1+\sqrt{B+1}\right)(\cos\left(2\pi\sqrt{B+1}\right)+\cosh
(2x\pi))}P_m(x^2)$$
and then integrate over $x$ from $0$ to $\infty$. Using the orthogonality
property of Wilson polynomials in (\ref{eb2}), we get the formula for $c_n$:
\begin{eqnarray}
c_n&=&\frac{4(-1)^n\pi^2(2n)!\,(2n+1)!}{(n!)^4\Gamma^2\left(n+1-\sqrt{B+1}
\right)\Gamma^2\left(n+1+\sqrt{B+1}\right)}\nonumber\\
&&\qquad\times\int_0^\infty dx\,\frac{x\e^{-x\pi}\tanh(x\pi)}{\cos\left(2\pi
\sqrt{B+1}\right)+\cosh(2 x \pi)}P_{n}(x^2).
\label{e59}
\end{eqnarray}

We conclude that we have decomposed the representation under which $\cP$
transforms into a direct sum of finite-dimensional irreducible representations.
Even though the operator $B$ is taken to be nonzero, the eigenvalues in the
difference-equation eigenvalue problem (\ref{e47}) remain unchanged. Thus, the
conclusion of Sec.~\ref{s4} that the parity operator $\cP$ transforms as the
direct sum of the spin-$0$ components of the finite-dimensional tensor
representations $(0,1)\oplus(0,3)\oplus(0,5)\oplus(0,7)\oplus\cdots$ remains
unchanged.

Finally, as we observed at the end of Sec.~\ref{s4}, we point out that the first
irreducible representation $f_0(B,W){\cal P}$ in the direct sum (\ref{e55}) is a
scalar. Thus, we may conclude that under a Lorentz transformation the parity
operator $\cP$ transforms as the operator $\e^{-i\pi\sqrt{B+W+1/4}}$.

\section{Final Remarks}
\label{s6}

The general case for (\ref{e23}) is obtained when all three parameters, $B$,
$M$, and $W$, are nonzero. For this case we cannot solve (\ref{e23}) in
terms of polynomials but we conjecture that the eigenvalues in (\ref{e28})
remain unchanged \cite{conj}. Assuming that this is indeed the case, we may
conclude that the Lorentz transformation properties of the parity operator
are unchanged from what we found in Secs.~\ref{s4} and \ref{s5}; namely, the
parity operator transforms like the direct sum of irreducible representations in
(\ref{e46}).

\section*{Acknowledgements}
We are greatly indebted to R.~Askey for identifying the polynomials in
(\ref{e29}) as Wilson polynomials. This work was supported in part by the
U.S.~Department of Energy.

\appendix
\section{Properties of Wilson Polynomials}
\label{a1}

In this appendix, we list the properties of the Wilson polynomials ${\rm W}_n(
x^2;a,b,c,d)$ with parameters $a=b=\half$, $c=d={\textstyle\frac{3}{2}}$
\cite{Askey,KS}.
For simplicity, we define the associated {\it monic} Wilson polynomials
\begin{equation}
P_n(x^2)\equiv\frac{(-1)^n(n+2)!}{(2n+2)!}{\rm W}_n\left(x^2;\half,\half,
{\textstyle\frac{3}{2}},{\textstyle\frac{3}{2}}\right).
\label{ea1}
\end{equation}
(The coefficient of the highest power in a {\it Monic} polynomial is unity.)
These polynomials are used in Sec.~\ref{s4}.

The orthogonality condition is
\begin{equation}
\frac{\pi^2}{4}\int_0^\infty dx\,x(1+4x^2)^2\frac{\sinh(x\pi)}{\cosh^3(x\pi)}P_n
(x^2)P_m(x^2)=\frac{(n!)^2[(n+1)!]^4[(n+2)!]^4}{[(2n+2)!]^2(2n+3)!}\delta_{nm}.
\label{ea2}
\end{equation}

The polynomials $P_n(x^2)$ satisfy a three-term recurrence relation:
\begin{equation}
P_n(x^2)=\left[x^2-\frac{1}{2}n(n+1)\right]P_{n-1}(x^2)+\frac{(n-1)^2 
n^2(n+1)^2}{4(2n-1)(2n+1)}P_{n-2}(x^2),
\label{ea3}
\end{equation}
where the first two polynomials are $P_0(x^2)=1$ and $P_1(x^2)=x^2-\frac{3}{4}$.

Some generating functions for these polynomials are
\begin{eqnarray}
\sum_{n=0}^\infty\frac{2(2n+2)!\,P_n(x^2)t^n}{(n!)^2[(n+2)!]^2}
&=& {}_2{\rm F}_1\left( \half+ix,\half+ix;1;-t\right) {}_2 {\rm F}_1\left( 
{\textstyle\frac{3}{2}}-ix,{\textstyle\frac{3}{2}}-ix;1;-t\right),\nonumber\\
\sum_{n=0}^\infty\frac{(2n+2)!\,P_n(x^2)t^n}{n![(n+1)!]^2(n+2)!}
&=& {}_2{\rm F}_1\left( \half+ix,{\textstyle\frac{3}{2}}+ix;2;-t\right)
{}_2{\rm F}_1\left( \half-ix,{\textstyle\frac{3}{2}}-ix;2;-t\right),\nonumber\\
\sum_{n=0}^\infty\frac{(2n+2)!\,P_n(x^2)t^n}{(n!)^2[(n+1)!]^2}
&=&\frac{1}{(1+t)^3} {}_4{\rm F}_3\left[ 
{\textstyle\frac{3}{2}},2,\half+ix,\half-ix;1,2,2;\frac{4t}{(1+t)^2}\right].
\label{ea4}
\end{eqnarray}

\section{\label{a2} Further Properties of Wilson Polynomials}

In this appendix, we list the properties of the Wilson polynomials ${\rm W}_n(
x^2;a,b,c,d)$ with parameters $a=b=\half$, $c=\half-\sqrt{B+1}$, $d=\half+
\sqrt{B+1}$ \cite{Askey,KS}. We define the associated monic polynomials
\begin{equation}
P_n(x^2)\equiv\frac{(-1)^nn!}{(2n)!}{\rm W}_n\left(x^2;\half,\half,\half-\sqrt{B
+1},\half+\sqrt{B+1}\right).
\label{eb1}
\end{equation}
These polynomials are used in Sec.~\ref{s5}.

Orthogonality:
\begin{eqnarray}
&&\quad4\pi^2\int_0^\infty dx\,x\frac{\tanh(x\pi)}{\cos\left(2\pi
\sqrt{B+1}\right)+\cosh(2 x\pi)} P_n(x^2) P_m(x^2)\nonumber\\
&=& \frac{(n!)^4 \Gamma^2\left(n+1-\sqrt{B+1}\right)
\Gamma^2\left(n+1+\sqrt{B+1}\right)}{(2n)!\,(2n+1)!} \delta_{nm}.
\label{eb2}
\end{eqnarray}

Recurrence relation:
\begin{equation}
P_n(x^2)=\left[x^2-\frac{B}{2}-\frac{1}{4}+\frac{n(n+1)}{2}\right]P_{n-1}(x^2)
+\frac{n^2[B-(n-1)(n+1)]^2}{4(2n-1)(2n+1)}P_{n-2}(x^2),
\label{eb3}
\end{equation}
with $P_0(x^2)=1$ and $P_1(x^2)=x^2+\half B+\quarter$.

Generating functions:
\begin{eqnarray}
&&\quad \sum_{n=0}^\infty\frac{(2n)!\,P_n(x^2)t^n}{(n!)^4}\nonumber\\
&&={}_2{\rm F}_1\left(\half+ix,\half+ix;1;-t\right){}_2
{\rm F}_1\left(\half-\sqrt{B+1}-ix,\half+\sqrt{B+1}-ix;1;-t\right)
\label{eb4}
\end{eqnarray}
and
\begin{eqnarray}
&& ~~\sum_{n=0}^\infty\frac{(2n)!\,P_n(x^2)t^n}{(n!)^2 
\Gamma\left(n+1-\sqrt{B+1}\right)\Gamma\left(n+1+\sqrt{B+1}\right)}\nonumber\\
&&\!\!\!\!\!\!\!\!\!\!=\frac{\sin\left(\pi\sqrt{B+1}\right)}{\pi\sqrt{B+1}}{}_2
{\rm F}_1\left(\half+ix,\half-\sqrt{B+1}+ix;1-\sqrt{B+1};-t\right)\nonumber\\
&&\qquad \qquad \qquad \times~{}_2{\rm F}_1\left(\half-ix, 
\half+\sqrt{B+1}-ix;1+\sqrt{B+1};-t\right)\nonumber\\
&&\!\!\!\!\!\!\!\!\!\!=\frac{\sin\left(\pi\sqrt{B+1}\right)}{\pi(1+t)\sqrt{B+1}}
{}_4{\rm F}_3\left(\half,1,\half+ix,\half-ix;1,1-\sqrt{B+1},1+\sqrt{B+1};
\frac{4t}{(1+t)^2}\right).
\label{eb5}
\end{eqnarray}


\begin{thebibliography}{99}
\bibitem{POS} For the irreducible representations of the Lorentz group if the
parameter $\ell_0=0$, then $(0,\ell_1)$ and $(0,-\ell_1)$ label exactly the
same representation. Thus, we may assume without loss of generality that
$\ell_1$ is a positive number. See Ref.~\cite{Gel}.
\bibitem{conj} Verifying this claim is difficult and requires heavy
numerical analysis. We believe that the situation is similar to
that for the eigenfunctions for the quantum harmonic and anharmonic oscillators.
In the former case, after the Gaussian is divided off, the eigenfunctions
satisfy the boundary conditions in a simple way; namely, by truncating in the
form of Hermite polynomials. In the latter case, after the Gaussian is removed,
the eigenfunctions do not truncate, but they still satisfy the boundary
conditions as one can verify numerically.
\end{thebibliography}
\end{document}